\documentclass[letterpaper, 10 pt, conference]{ieeeconf}  

\IEEEoverridecommandlockouts                              

\usepackage{amsmath} 
\allowdisplaybreaks 
\usepackage{amssymb}  
\usepackage{graphicx, caption, subcaption}
\usepackage{algorithm}
\usepackage{algorithmic}
\let\labelindent\relax
\usepackage{enumitem}

\newcommand{\R}{\mathbb{R}}
\newcommand{\cK}{\mathcal{K}}
\newcommand{\N}{\mathbb{N}}

\newcommand{\br}[1]{\operatorname{br}_{#1}}
\newcommand{\sgn}{\text{sgn}}
\newcommand{\NE}{{NE}}

\newcommand{\eqdef}{:=}

\newcommand{\ie}{\textit{i.e.} }

\newcommand{\cf}{\textit{cf.} }

\usepackage{mathrsfs}

\newtheorem{theorem}{Theorem}

\newtheorem{lemma}{Lemma}
\newtheorem{definition}{Definition}
\newtheorem{assumption}[theorem]{Assumption}
\newtheorem{remark}[theorem]{Remark}

\title{\LARGE \bf
Nash equilibria in scalar discrete-time linear quadratic games
}

\author{Giulio Salizzoni, Reda Ouhamma, and Maryam Kamgarpour
\thanks{G. Salizzoni, R. Ouhamma and M. Kamgarpour are with SYCAMORE Lab, School of Engineering, EPFL, Lausanne, Switzerland (email: {\tt\small giulio.salizzoni@epfl.ch}, {\tt\small reda.ouhamma@epfl.ch},{\tt\small maryam.kamgarpour@epfl.ch}) }
\thanks{This research is gratefully supported by the Swiss National Science Foundation, grant number 207984 and by the SNSF NCCR Automation Grant.}%
}

\begin{document}

\maketitle
\thispagestyle{empty}
\pagestyle{empty}

\begin{abstract}
An open problem in linear quadratic (LQ) games has been characterizing the Nash equilibria. This problem has renewed relevance given the surge of work on understanding the convergence of learning algorithms in dynamic games. This paper investigates scalar discrete-time infinite-horizon LQ games with two agents. Even in this arguably simple setting, there are no results for finding \textit{all} Nash equilibria. By analyzing the best response map, we formulate a polynomial system of equations characterizing the linear feedback Nash equilibria. This enables us to bring in tools from algebraic geometry, particularly the Gröbner basis, to study the roots of this polynomial system. Consequently, we can not only compute all Nash equilibria numerically, but we can also characterize their number with explicit conditions. For instance, we prove that the LQ games under consideration admit at most three Nash equilibria. We further provide sufficient conditions for the existence of at most two Nash equilibria and sufficient conditions for the uniqueness of the Nash equilibrium. Our numerical experiments demonstrate the tightness of our bounds and showcase the increased complexity in settings with more than two agents.
\end{abstract}

\section{INTRODUCTION}

Game theory is the study of systems of multiple interacting agents, each with her own goals, constraints, and decision-making capabilities. This framework of interaction is popular and has been successful for different real-world applications such as robotics \cite{cappello2021distributed}, autonomous driving \cite{fisac2019hierarchical, ren2025chance} and electricity markets \cite{paccagnan2016aggregative, kirsch2025distributed}. 
The control of such systems presents unique challenges due to the complex dynamics of interactions among agents. More generally, in non-cooperative games, the agents are interested in minimizing their cost function and cannot collaborate or communicate. A common objective in these settings is finding a Nash equilibrium, which is a set of policies such that no agent can improve her cost by unilaterally changing her policy. 

In this paper, we focus on linear quadratic (LQ) games, a benchmark setup of games with continuous states and actions. LQ games are the multi-agent extension of the LQ problem, a popular problem in control theory characterized by linear system dynamics and a quadratic cost function. The LQ problem is known to have a unique optimal linear feedback controller known as the linear-quadratic regulator (LQR), which is obtained by solving the so-called algebraic Riccati equation. On the other hand, LQ games entail more challenging Riccati equations which relate the agents in a coupled fashion and may admit many solutions. In addition, a key objective in LQ games is the characterization of all Nash equilibria. The latter stems from the practical goal of designing optimal controllers for specific applications. For instance, a critical aspect in practical scenarios is designing the cost matrices $Q$ and $R$ which govern the trade-off between state regulation and control effort \cite{anderson2007optimal}. Since LQ games potentially have many Nash equilibria, which may offer different trade-offs in terms of stability and efficiency, it is crucial to compute all Nash equilibria of an LQ game for an effective design of the cost matrices.

Computing the Nash equilibria in the differential setting (also called time-continuous) has received considerable attention. For instance, \cite{lukes1971global} provides conditions on the cost functions and the system dynamics for the uniqueness of the Nash equilibrium in the finite horizon case. In the infinite horizon case, \cite{Papav} provides sufficient conditions for the existence of linear feedback Nash equilibria in two-agent games. In the same setting, \cite{Gajic} claims that the coupled algebraic Riccati equations have a unique Nash equilibrium under conditions of detectability and stabilizability of the system. However, \cite{EngwerdaInfHorDif} and \cite{EngwerdaNE} disprove this result by demonstrating that in the scalar case the Nash equilibria are either one or three. For the non-scalar case, \cite{engwerda2005} proves that there could be either zero, one, or multiple solutions.

Discrete-time LQ games have proven to be more challenging for the purpose of characterizing Nash equilibria. For instance, existing results in the differential setting leverage the fact that the optimal parameters (Nash equilibria) depend linearly on the solutions of the Riccati equations. In the discrete-time case, this linearity does not hold because of additional product terms leading to coupled Riccati equations. Consequently, existing analyses do not characterize Nash equilibria as well as in the differential case and are limited to proving existence or uniqueness under special conditions. For the finite-horizon discrete-time case, \cite{BasarFiniteHorizon} provides necessary and sufficient conditions for the uniqueness of the Nash equilibrium, and \cite{PotentialDecoupledLQ} derives conditions for the existence of Nash equilibria if the agents' dynamics are decoupled. For the infinite-horizon case, \cite{Basar98} highlights that a linear feedback Nash equilibrium can be obtained by taking the limit of the Nash equilibrium of the time-truncated version of the game provided that this limit exists. However, as argued in \cite[Proposition 6.3]{Basar98}, providing conditions for the existence of this limit is challenging. More broadly, the conditions for the existence and the number of a Nash equilibrium in the infinite-horizon discrete-time setting are still not well understood. 

In this paper, we study infinite-horizon discrete-time scalar two-player LQ games as a first step toward understanding Nash equilibria in the non-scalar case with more agents. 

Concurrent with our work, \cite{nortmann2023feedback} investigate infinite horizon discrete time scalar two-players LQ games, deriving explicit conditions for the existence and number of Nash equilibria. The authors employ a graphical analysis of plane curves associated with the coupled Riccati equations. Their results are presented in terms of solutions of algebraic functions involving square roots of polynomials. While the equations are analytically complex, and numerical methods may face challenges due to the presence of square roots, their results provide the first work characterizing the existence and cardinality of equilibria. However, it is still not clear how to compute the Nash equilibria explicitly. They later extended the results to scalar $N$-players setting in \cite{nortmann2025feedback}.

\textbf{Contribution:} This paper is concerned with computing all linear feedback Nash equilibria for scalar discrete-time infinite-horizon LQ games with two agents. We leverage the best response maps to demonstrate that computing Nash equilibria is equivalent to solving a system of coupled polynomial equations. By studying this system using algebraic geometry techniques, we can extend the literature on scalar two-player LQ games as follows:
\begin{itemize}
    \item Using the notion of Gröbner bases from algebraic geometry, we provide a univariate polynomial whose real roots characterize the Nash equilibria; 
    \item The obtained polynomial expression allows us to: \begin{itemize}
        \item Prove that the Nash equilibria are at most three;
        \item Provide explicit sufficient conditions for the uniqueness of the Nash equilibrium and sufficient conditions for the existence of at most two Nash equilibria;
        \item Compute all Nash equilibria using root-finding algorithms for polynomials.
    \end{itemize}
    \item We validate our bounds by presenting examples of games with one, two, and three Nash equilibria. 
\end{itemize}

\section{PROBLEM STATEMENT}\label{sec:problemStatement}

\textbf{Notations.} We denote by $\R, \N$ the sets of real and natural numbers, respectively. For $n \in \N$, we denote by $[n]$ the set $\{1,2,\dots,n\}$.

\textbf{Setting.} In this paper, we consider scalar two-player linear quadratic games, a natural extension of the linear quadratic regulator (LQR) problem. The dynamics of the game are:
\begin{align*}
    x_{t+1} = a x_t + \sum_{i=1}^2 b_i u_{i,t},
\end{align*}
where $x_t \in \R$ is the state of the system at time $t$, $u_{i,t} \in \R$ is the control parameter of agent $i$ at time $t$, $a \in \R$ and $b_i \in \R$, $b_i \neq 0$ for all $i \in [2]$. We will use the index $i\in [2]$  to denote player $i$, and $j$ for the other one.
We restrict our attention to the class of linear feedback policies defined below.
\begin{assumption}{(linear feedback)}
    The agents use a linear state feedback policy, \ie $u_{i,t} = -k_i x_t$ where $k_i \in \R$ is the control parameter of agent $i$.
\end{assumption}

A common assumption for LQ games is the stabilizability of the system, meaning the existence of a pair $(k_1,k_2) \in \R^2$ such that $\left| a - b_1 k_1 - b_2 k_2 \right| < 1$. In scalar LQ games, this condition is always met. In fact, for any $a$ and any $k_1$, we can choose $k_2$ equal to $\frac{a - b_1 k_1}{b_2}$ obtaining $ a - b_1 k_1 - b_2 k_2 = 0$.

We consider an LQ game in which each agent $i$ aims to minimize the following cost:
\begin{align*}
    J_i(x_0,k_i,k_{j}) &= \sum_{t = 0}^\infty \left(q_i x_t^2 + r_i u_{i,t}^2 \right) = \sum_{t = 0}^\infty \left(q_i + r_i k_i^2 \right) x_t^2 ,
\end{align*}
with parameters $r_i , q_i > 0$. 

In LQ games, each agent seeks to minimize her cost function, and the objective is reaching a Nash equilibrium. 
\begin{definition}
    A pair of controls $k^\NE \in \R^2$ is a Nash equilibrium if
\begin{align*}
    J_i(x_0,k_i^\NE,k_{j}^\NE) \leq J_i(x_0,k_i,k_{j}^\NE), \; \forall k_i \in \R, \forall i \in [2]. 
\end{align*}
\end{definition} 
When a Nash equilibrium is reached, no agent can decrease her cost function by unilaterally changing her policy. The existence of a linear feedback Nash equilibrium in LQ games depends on the solvability of a set of coupled algebraic Riccati equations. While it is possible to give a precise condition for the unique solvability of the aforementioned set in the finite horizon case (\cf Remark 6.5 in \cite{Basar98}), this has not been done for the infinite horizon case (\cf remark 6.16 in \cite{Basar98}).

Notice that in \cite{nortmann2023feedback} the authors also examine the case where $q_i = 0$, but they need to require admissibility—meaning that the Nash equilibrium policies must stabilize the system. In contrast, we chose to consider $q_i > 0$, because this ensures that the Nash equilibrium policies are always stabilizing. If they were not stabilizing, the cost would become infinite, which would exceed the cost an agent could achieve by selecting $k_i = \frac{a - b_{j} k_{j}}{b_i}$, making it impossible for such a policy to be a Nash equilibrium. While the value of $q_i$ can be chosen arbitrarily small and would have minimal impact on the results, it allows us to avoid certain limiting cases that are beyond the scope of interest. The case where $r_i = 0$ is discarded as it is trivial. In fact, if $r_i = 0$, the optimal control parameter for agent $i$ is to choose $k_i = \frac{a - b_{j} k_{j}}{b_i}$.

\begin{assumption}
    The state coefficient $a$ is strictly positive.
\end{assumption}
This assumption does not lead to any loss of generality. To understand why, consider the closed loop of the dynamic, $ a_{cl} = a - b_1 k_1 - b_2 k_2$ , which can be written as
\begin{align} \label{eq:costacl}
    a_{cl} = \left( |a| - \sum_{i=1}^2 b_i |k_i| \frac{\sgn(k_i)}{\sgn(a)} \right) \sgn(a).
\end{align}
In the cost function $a_{cl}$ appears squared,
\begin{align*}
    J_i(x_0,k_i,k_{j}) = \sum_{t = 0}^\infty \left(q_i + r_i |k_i|^2 \right) (a_{cl}^2)^t x_0^2.
\end{align*}
Thus, for any Nash equilibrium $(k_1^{\NE},k_2^{\NE})$ of a game with $a \in \R_+ $, the pair $(-k_1^{\NE},-k_2^{\NE})$ is a Nash equilibrium of the corresponding game with $-a$, and vice versa. Moreover, if $a = 0$, the problem becomes trivial, as the unique Nash equilibrium is $(0,0)$.

In addition, whenever $b_1$ and $b_2$ are not equal to $1$, we can study an equivalent problem where we substitute $b_i k_i$ with $\tilde{k}_i$ and $\frac{r_i}{b_i^2}$ with $\tilde{r_i}$. The closed loop would then be
\begin{align} \label{eq:dynamic}
    a_{cl} = a - \tilde{k}_1 - \tilde{k}_2 ,
\end{align}
and the cost function
\begin{align} \label{eq:costFunction}
    J_i(x_0,k_i,k_{j}) =&  \sum_{t = 0}^\infty \left(q_i + \tilde{r}_i \tilde{k}_i^2 \right) a_{cl}^2 x_0^2.
\end{align}

\section{COMPUTING THE NASH EQUILIBRIA}

This section presents the necessary steps to compute the Nash equilibria. The first subsection defines the system we must solve to calculate the Nash equilibria. In the second one, we present the concept of Gröbner bases, which allows us to solve the system.

\subsection{A system characterizing Nash equilibria}

In the following lemma, we provide a system of equations characterizing the Nash equilibria of the game.
\begin{lemma} \label{lemma:system}
    A pair of policies $(k_1,k_2)$ is a Nash equilibrium of the game with dynamics \eqref{eq:dynamic} and cost functions \eqref{eq:costFunction} if and only if it is a solution of the following system
    \begin{equation}\label{sys:derivatives}
        \begin{cases}
            \!(a\!-\!k_2)r_1 k_1^2 \!+\! (r_1 \!+\! q_1\! - \!(a\!-\!k_2)^2 r_1 )k_1 \!- \!(a\!-\!k_2)q_1\!=\! 0, \\[10pt]
            \!(a\!-\!k_1)r_2 k_2^2 \!+\! (r_2 \!+\! q_2\! - \!(a\!-\!k_1)^2 r_2 )k_2 \!- \!(a\!-\!k_1)q_2\!=\! 0, \\[10pt]
            |a - k_1 - k_2| < 1,
        \end{cases} \tag{P}
    \end{equation}

\end{lemma}
\begin{proof}
    Given a fixed policy $k_{j}$ of the opponent, agent $i$ is facing an LQR problem with dynamics $x_{t+1} = (a - k_{j}) x_t + u_{i,t},$ and cost $\sum_{t = 0}^\infty \left(q_i x_t^2 + r_i u_{i,t}^2 \right)$. Thus, thanks to the gradient dominance property of the LQR \cite{Fazel18}, we know that a pair $(k_1,k_2) \in \R^2$ is a Nash equilibrium if and only if the derivative of each agent's cost function is zero:
\begin{align}\label{sys:derivCost}
\frac{\partial J_1 (x_0, k_1, k_2)}{\partial k_1} = \frac{\partial J_2 (x_0, k_1, k_2)}{\partial k_2} = 0  \tag{D}
\end{align}
    
    Consider $(k_1,k_2) \in \cK$, where $\cK \eqdef \{(x,y) \in \R^2 \; : \; |a - x - y|<1\}$ is the set of stabilizing policies.  The cost function of agent $i$ can be written as:
\begin{align} \label{eq:costFraction}
    J_i(x_0,k_1,k_2) =& \sum_{t = 0}^\infty \left(q_i + r_i k_i^2 \right) a_{cl}^{2t} x_0^2 = \frac{  q_i + r_i k_i^2 }{ 1 - a_{cl}^2 } x_0^2.
\end{align}
This formulation of the cost is valid only in the set of stabilizing policies, otherwise the cost would not be finite. The derivative is equal to
\begin{align*}
    \frac{\partial J_i (x_0, k_1, k_2)}{\partial k_i} = 2 \frac{ r_i k_i (1 - a_{cl}^2 ) - ( q_i + r_i k_i^2 ) a_{cl}  }{ \left( 1 - a_{cl}^2 \right)^2 } x_0^2 = 0.
\end{align*}
After dividing everything by $2x_0^2$ , multiplying everything by $\left( 1 - a_{cl}^2 \right)^2$, and substituting $a_{cl}$ with $a - k_1 - k_2$ we have
\begin{align*}
    &r_i k_i - ( a - k_{j} )^2 r_i  k_i + 2 ( a - k_{j} ) r_i  k_i^2 - r_i k_i^3 \\
    &- ( a - k_{j} ) q_i + q_i k_i - ( a - k_{j} ) r_i  k_i^2  + r_i k_i^3 = 0,
\end{align*}
from which we obtain
\begin{align} \label{eq:brlong}
    \!( a \!-\! k_{j} ) r_i k_i^2 \!+\! ( r_i \!+ \!q_i \!- \!( a \!- \!k_{j} )^2 r_i ) k_i \!-\! ( a \!-\! k_{j} ) q_i \! =\! 0.
\end{align}
One of the two solutions of \eqref{eq:brlong} does not respect the condition $|a - k_1 - k_2|<1$. Thus, \eqref{eq:brlong} has a unique admissible solution.
By substituting \eqref{eq:brlong} in the system \eqref{sys:derivCost} and adding the condition $|a - k_1 - k_2|<1$,we obtain system \eqref{sys:derivatives}.
\end{proof}
The sought-after Nash equilibria are strictly included in the common real roots of the above polynomials: namely, they need to ensure stability. From B\'ezout's theorem, the number of solutions of a system of multivariate polynomials is the product of the degrees of the polynomials. Therefore, the system \ref{sys:derivatives} possesses at most nine solutions, and the number of Nash equilibria is at most nine. However, this upper bound on the number of solutions also accounts for complex solutions, solutions outside $\cK$, and solutions with multiplicity. In Section \eqref{sec:characNE}, we characterize the number of Nash equilibria, while in the next subsection, we present a tool to compute all the real roots.

\begin{remark}
The system \eqref{sys:derivatives} can also be derived from the coupled algebraic Riccati equations and the corresponding optimal control equations. However, transforming the latter equations into a polynomial form is more tedious. Instead, it is straightforward to derive the system of interest directly from the cost functions.
\end{remark}

\subsection{A simpler system using Gröbner Basis}

To tackle the complexity of System \eqref{sys:derivatives}, we introduce the concept of Gröbner bases from algebraic geometry. Given an initial system of multivariate polynomials, a Gröbner basis is a simpler system of polynomials with the same solutions as the original one. Gröbner bases are the main tool for solving complicated systems of polynomials and characterizing their solutions (\cf \cite{sturmfels2005grobner}). 

The main algorithm for finding Gröbner bases is called Buchberger's algorithm \cite{buchberger1970algorithmisches,buchberger1965algorithmus,buchberger1985grobner}. While explaining such an algorithm is out of the scope of this paper, we mention that it is essentially a generalization of Gauss elimination for systems of polynomials, \ie it obtains an equivalent system by performing basic operations like addition and multiplication. We computed a Gröbner basis for System \eqref{sys:derivatives} using the computer software Macaulay2\footnote{Macaulay2 is a software system designed for symbolic algebraic computations, including finding Gröbner bases.} with a Lexicographical ordering (lex). The new (equivalent) system is comprised of ten equations, nine of which are coupled and at least as complicated as the original system, and the tenth equation is $g(k_2)=0$ where:

\vspace{-0.3cm}
\begin{align}\label{eq:Grobner}
    g(k_2) =& a r_1^2 r_2^2 k_2^5 \nonumber\\
    &+(-2.5a^2 r_1^2r_2^2+q_2r_1^2r_2-q_1r_1r_2^2+r_1^2r_2^2)k_2^4 \nonumber\\
    &+(2a^3 r_1^2 r_2^2 - 2aq_2 r_1^2 r_2 + 2aq_1 r_1r_2^2 - 2ar_1^2 r_2^2)k_2^3 \nonumber\\
    &+(-.5 a^4 r_1^2 r_2^2+a^2q_2r_1^2r_2-a^2 q_1r_1r_2^2+a^2r_1^2 r_2^2 \nonumber \\
    &\quad +.5q_2^2r_1^2-.5q_1^2r_2^2 -q_1r_1r_2^2 -.5r_1^2r_2^2)k_2^2 \nonumber \\
    &-aq_2^2r_1^2 k_2+.5a^2q_2^2r_1^2 \tag{G}
\end{align}

\noindent
Observe that $g$ is a fifth-degree polynomial that is uncoupled, \ie only depends on the control of the second player. Moreover, by the properties of a Gröbner basis, the solutions of \eqref{sys:derivatives} are exactly the solutions of the new system containing the ten equations that are in the stable set $\cK$. Since equation \eqref{eq:Grobner} is part of the new system, the roots of equation \eqref{eq:Grobner} contain those of the system. Consequently, using any root-finding algorithm on the polynomial \eqref{eq:Grobner}, we can find all Nash equilibria of the underlying game, this is the first result of this kind for discrete-time infinite-horizon LQ games.

\begin{remark}
    Macaulay2 is a symbolic solver, \ie does not resort to approximations for computing a Gröbner basis. The new system is equivalent to \eqref{sys:derivatives}.
\end{remark}

\section{CHARACTERIZING THE NUMBER OF NASH EQUILIBRIA} \label{sec:characNE}

In this section, we use the results from the previous section to characterize the number of Nash equilibria.

\subsection{Main result}

Let us denote by $\Delta_g$ the discriminant of the polynomial $g$ of equation \eqref{eq:Grobner}. The discriminant of a polynomial $A(x)$ with leading monomial $a_n x^n$ is defined as $(-1)^{n(n-1)/2} / a_n$ times the determinant of the Sylvester matrix of $A$ and $A'$. The Sylvester matrix of $A$ and $A'$ is an $(2n-1)*(2n-1)$ matrix defined by basic operations on the coefficients of $A$ and $A'$. See \cite{woody2016polynomial} for definitions and a complete exposition. Importantly, the discriminant entails interesting properties about the roots of a polynomial.


We are now ready to present our main result.
\begin{theorem}\label{thm:main}
    Given a two-player scalar linear quadratic game, denote by $\Delta_g$ the discriminant of the polynomial $g$ from equation \eqref{eq:Grobner}. Then, the following statements hold:
    \begin{enumerate}
        \item There exists a Nash equilibrium;
        \item There are at most three Nash equilibria;
        \item If $\Delta_g=0$, then there are at most two Nash equilibria;
        \item If $\Delta_g<0$, then the Nash equilibrium is unique.
    \end{enumerate}
\end{theorem}

\vspace{0.1cm}

The proof is provided in the next section. We first connect our results with existing ones. In the differential setting, \cite{EngwerdaInfHorDif} proves that the Nash equilibria are either one or three. On the contrary, our results do not exclude having two Nash equilibria in discrete-time settings. Subsequently, \cite{NEPossieri} leverages Gröbner bases to simplify the system of Riccati equations in the differential setting, and shows that the Nash equilibria are at most three. However, their approach relies on the linear dependence of the optimal control on the solution of the Riccati equation in the differential setting, which becomes highly non-linear in our discrete-time setting. 

Compared to \cite{nortmann2023feedback}, whose approach requires solving algebraic equations to determine the number of Nash equilibria, we provide an exact result for the number of equilibria and we can compute all Nash equilibria efficiently. The Nash equilibria are a subset of the roots of the fifth-degree polynomial $g$ that are in the stable region for the system. The discriminant $\Delta_g$ can be evaluated to determine the number of Nash equilibria without numerical approximations, and standard numerical solvers can efficiently identify the roots of $g$, in contrast to the non-trivial challenge of computing the roots for the algebraic functions of \cite[Lemma 2]{nortmann2023feedback}.

\begin{remark}
    The system \eqref{sys:derivatives} can be generalized to LQ games with $n \ge 2$ agents. The corresponding system characterizing the best response maps contains $n$ third-degree polynomials in $n$ variables. By applying the theorem of B\'ezout we know that there are $3^n$ solutions to this system, some of which are complex or possibly outside the optimal interval (Lemma \ref{lem:OptimalRange}). From \cite{nortmann2025feedback} we know that the game can have at most $2^n - 1$ Nash equilibria. We attempted using Macaulay2 to find Gröbner bases for the system three-agent LQ games, but the solver did not seem to converge. We believe that the latter is due to the exponentially increasing complexity of Buchberger's algorithm, see \cite{dube1990structure} and related works.
\end{remark}


\subsection{Proof of Theorem \ref{thm:main}}

This section presents the proof of Theorem \ref{thm:main}. We first present an essential technical Lemma with its proof and then move to the proof of Theorem \ref{thm:main}.

\paragraph{Best response} We present the following lemma which provides an explicit form of the best-response functions and their properties.

\begin{lemma} \label{lem:OptimalRange}
    Denote by $\br{i}$ the best response function of agent $i$,\footnote{The best response map $\br{i}$ takes as input a control policy $k_{j}$ of agent $j$ and returns the optimal control parameter for agent $i$.} then for all $k_{j}$:  
    \begin{equation} \label{eq:br}
        \br{i}(k_{j})\! =\! (a \!-\! k_{j}) \frac{ ( ( a \!\!-\!\! k_{j} )^2 \!-\! 1) r_i \!+\! q_i\! +\! \sqrt{\! s_i( k_{j} ) } }{ ( ( a \!\!-\!\! k_{j} )^2 \!+\! 1) r_i \!+\! q_i \!+\! \sqrt{\! s_i( k_{j} ) } },
    \end{equation}
    where $s_i( k_{j} ) = \left( ( ( a - k_{j} )^2 - 1) r_i + q_i \right)^2 + 4 q_i r_i$.
    \vspace{0.2cm}
                
    Consider a control parameter $k_{j}$, it holds that:
    \begin{enumerate}
        \item $\sgn \left( \br{i}(k_{j}) \right) = \sgn (a - k_{j})$,
        \item $\left| \br{i}(k_{j}) \right| \leq \left| a - k_{j} \right|$, with equality if and only if $k_{j} = a$.
    \end{enumerate}
    In addition, for a Nash equilibrium $(k_1^\NE ,k_2^\NE) \in \R^2$ the following bounds hold:
    \begin{enumerate}[resume]
        \item $0< k_i^\NE < a - k_{j}^\NE < a, \quad \text{for } i \in [2].$
    \end{enumerate}
\end{lemma}

\vspace{0.2cm}

\begin{proof} \underline{Best-response function:} In our setting, thanks to the properties of scalars, the algebraic Riccati equation becomes
\begin{align*}
    p_i = q_i + (a -k_{j})^2 p_i - \frac{(a -k_{j})^2 p_i^2}{ r_i + p_i }.
\end{align*}
By multiplying everything by $r_i + p_i$, we obtain a second-degree equation in $p_i$. One of the two solutions is negative, while $p_i$ must be positive. The unique positive solution is
\begin{align} \label{eq:pi}
    p_i = \frac{1}{2}\! \left( \!(\!a \!-\!k_{j}\!)^2 \!r_i \!+\! q_i \!+ \sqrt{ \left( \!(\!a \!-\!k_{j}\!)^2 \!r_i\! + \!q_i\! \right)^2 \!+ \!4 r_i q_i } \!\right).
\end{align}
The best response of agent $i$ is $k_i = (a -k_{j}) \frac{ p_i }{ r_i + p_i}$.
By substituting $p_i$ with \eqref{eq:pi}, it is possible to recover \eqref{eq:br}.

\underline{Proof of statements 1) and 2):} The term
\begin{align*}
     \frac{ ( ( a - k_{j} )^2 - 1) r_i + q_i + \sqrt{ s_i( k_{j} ) } }{ ( ( a + k_{j} )^2 + 1) r_i + q_i + \sqrt{ s_i( k_{j} ) } }
\end{align*}
is always positive, since $\sqrt{ s_i( k_{j} ) } > | ( ( a - k_{j} )^2 - 1) r_i + q_i |$, and smaller than $1$. Thus, the best response has the same sign of $a - k_{j}$ and is smaller in absolute value.

\underline{Proof of statement 3):} Consider now a Nash equilibrium pair $(k_1^\NE,k_2^\NE)$ and suppose by contradiction that $k_1^\NE < 0$. Since at a Nash equilibrium, each policy is the best response of the other one, we can use the first and second statements of Lemma \ref{lem:OptimalRange}. Therefore, we have $\sgn(k_1^\NE) = \sgn(a - k_2^\NE) = -1$, which implies that $k_2^\NE > a$. Then:
\begin{equation*}
    |k_1^\NE| < |a-k_2^\NE| = - a + k_2^\NE, \: \text{ and } k_2^\NE < a - k_1^\NE.
\end{equation*}
The first inequality implies that $a < k_1^\NE + k_2^\NE$, while the second one entails that $k_1^\NE + k_2^\NE < a$. This is a contradiction, therefore $\sgn(k_i^\NE) = \sgn (k_{j}^\NE) = 1$. 
By using Statements 1 and 2 of Lemma \ref{lem:OptimalRange} a second time we find that:
\begin{equation*}
    0 < k_i^{\NE} < a - k_{j}^{\NE} < a, \: \forall i \in [2]
\end{equation*}
which concludes the proof.
\end{proof}

\vspace{0.2cm}

\begin{remark}
    We emphasize that assuming $a >0$ is without loss of generality thanks to the quadratic dependence on $k_i$ and $a_{cl}$ in equation \ref{eq:costacl}. For Lemma \ref{lem:OptimalRange}, the proof can be adapted for $a < 0$. Statements 1 and 2 still hold, and the third becomes $a < a - k_{j}^\NE < k_i^\NE < 0, \quad \text{for } i,j \in [2], \; i \neq j.$ 
\end{remark}

\paragraph{Proof of Theorem \ref{thm:main}} We now provide the proof of Theorem \ref{thm:main}.

\begin{proof} \underline{Proof of statement 1):} Here, we prove the existence of a Nash equilibrium, leveraging Lemma \ref{lem:OptimalRange} and the best response equation \eqref{eq:br}. Observe that in two-player settings, a pair of control parameters $(k_1,k_2) \in \R^2$ is a Nash equilibrium if and only if: $k_i = \br{i} \circ \br{j} (k_i)$, \ie both $k_1$ and $k_2$ are fixed points of the respective composition of best response maps. Thanks to Lemma \ref{lem:OptimalRange}, we know that the solution should be in the range $(0, a)$. Therefore, our strategy to prove the existence of a Nash equilibrium is to find a zero of the function $h: x \in \R \mapsto \br{1} \circ \br{2}(x) - x$ in the interval $(0, a)$.

\vspace{0.2cm}

\textbf{$h(0)$ is positive:} From Lemma \ref{lem:OptimalRange} we have that $|\br{2}(0)| < |a|$ and that $\sgn( \br{1}(\br{2}(0)) = \sgn(a - \br{2}(0))$ which is positive. Hence $h(0) = \br{1}(\br{2}(0))$ is positive.

\vspace{0.2cm}

\textbf{$h(a)$ is negative:} Using again \eqref{eq:br}, we can verify that $\br{2}(a) = 0$. Therefore, we have $\br{1}(\br{2}(a)) < a$ using Lemma \ref{lem:OptimalRange}, \ie $h(a) = \br{1}(\br{2}(a)) - a$ is negative.

\vspace{0.1cm}

We have proved that $h(0)>0$ and $h(a)<0$, and the best response maps are continuous, then, $h$ is continuous as the composition of continuous functions. Therefore, using the mean value theorem, we deduce the existence of $k_1^* \in (0,a)$ such that $h(k_1^*)=0$. Consequently, there exists a Nash equilibrium for the game.

\underline{Proof of statement 2):} Here, we bound the number of Nash equilibria based on the Gröbner basis \eqref{eq:Grobner} and Lemma \ref{lem:OptimalRange}. Recall that the Nash equilibria are a subset of the roots of the polynomial $g$.
    
Since $g$ is a univariate fifth-degree polynomial with real coefficients, it admits at most five real roots. Additionally, we have $\lim_{k_2 \rightarrow-\infty} g(k_2) = -\infty$ and $g(0) = .5a^2q_2^2r_1^2 >0$. Thus, the continuity of $g$ implies the existence of a root in the interval $(-\infty,0)$. Similarly, $g(a) = -.5 a^2 q_1^2 r_2^2 - a^2 q_1 r_1 r_2^2 - .5 a^2 r_1^2 r_2^2 <0$ and $\lim_{+\infty} g(k_2) = +\infty$, which implies the existence of a root of $g$ in the interval $(0,+\infty)$. By Lemma \ref{lem:OptimalRange}, we know that the Nash equilibria are in the interval $(0, a)$. Consequently, there are at most three Nash equilibria for scalar two-player discrete-time LQ games.

\underline{Proof of statements 3) and 4):} Here, we provide sufficient conditions for the existence of a maximum of one or two Nash equilibria. The discriminant of $g$, $\Delta_g$ is a polynomial function of the system parameters, and it can be used to characterize the number of roots of $g$ using the number of roots theorem  \cite{nickalls1996geometry}.

\textbf{$\Delta_g = 0$:} Here, at least one root of $g$ has multiplicity greater than one, see \cite{nickalls1996geometry}. Therefore, from part 2 of the theorem, we conclude the maximum number of distinct roots in $(0, a)$ is two. This proves Statement 3.

\textbf{$\Delta_g < 0$:} Here, the number of real roots of $g$ is congruent to three modulo four, see \cite{nickalls1996geometry}. Since the number of real roots varies from one to five, it must be three. Additionally, as shown in the proof of part 2, two real roots are outside $(0, a)$, therefore there is exactly one solution in $(0, a)$. This proves the Statement 4.
\end{proof}

\begin{remark}
    Theorem \ref{thm:main} cannot readily be generalized beyond the two-agent scalar case. In particular, note that the algebraic geometry tool alone was not sufficient to get the tight bounds and in all the statement proofs, we used Lemma \ref{lem:OptimalRange}, which is specific to two-agent scalar LQ games and cannot be extended to more general LQ games. 
\end{remark}

\section{NUMERICAL EXAMPLES}

\begin{figure}[!h]
    \centering
    \begin{subfigure}{\linewidth}
        \centering
        \includegraphics[width=.88\linewidth]{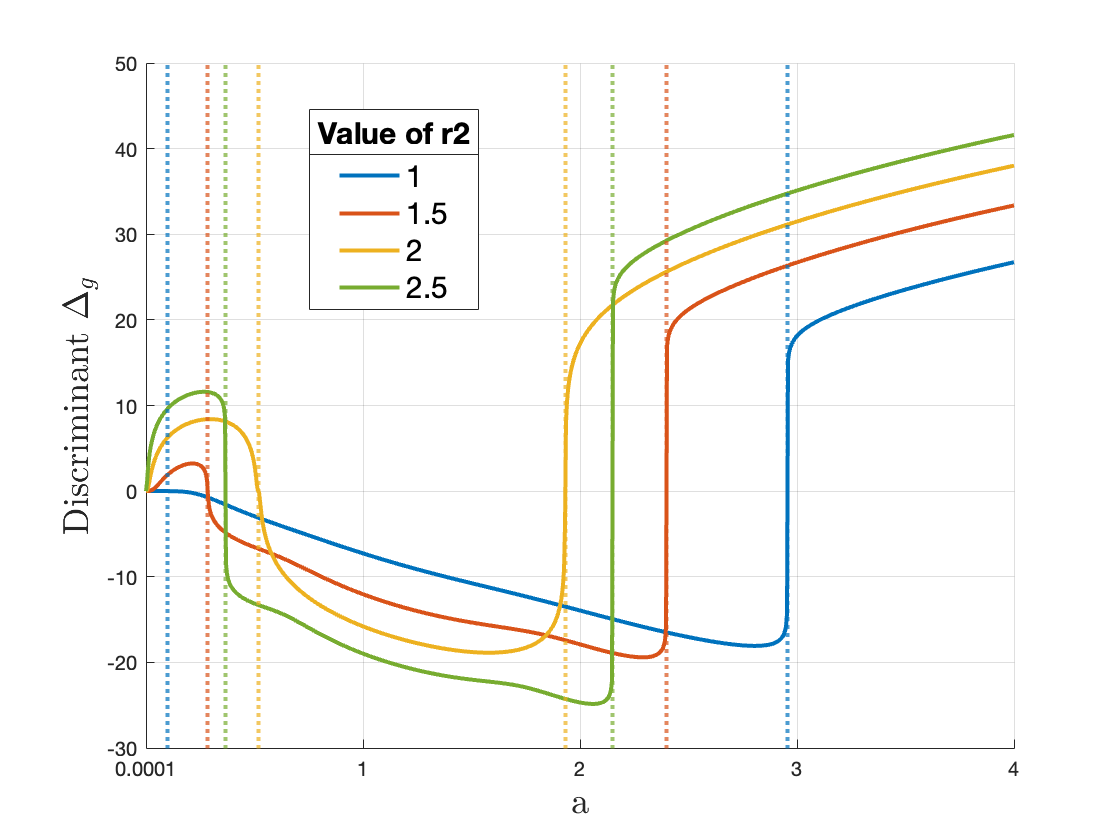}
    \end{subfigure}
    \begin{subfigure}{\linewidth}
        \centering
        \includegraphics[width=.88\linewidth]{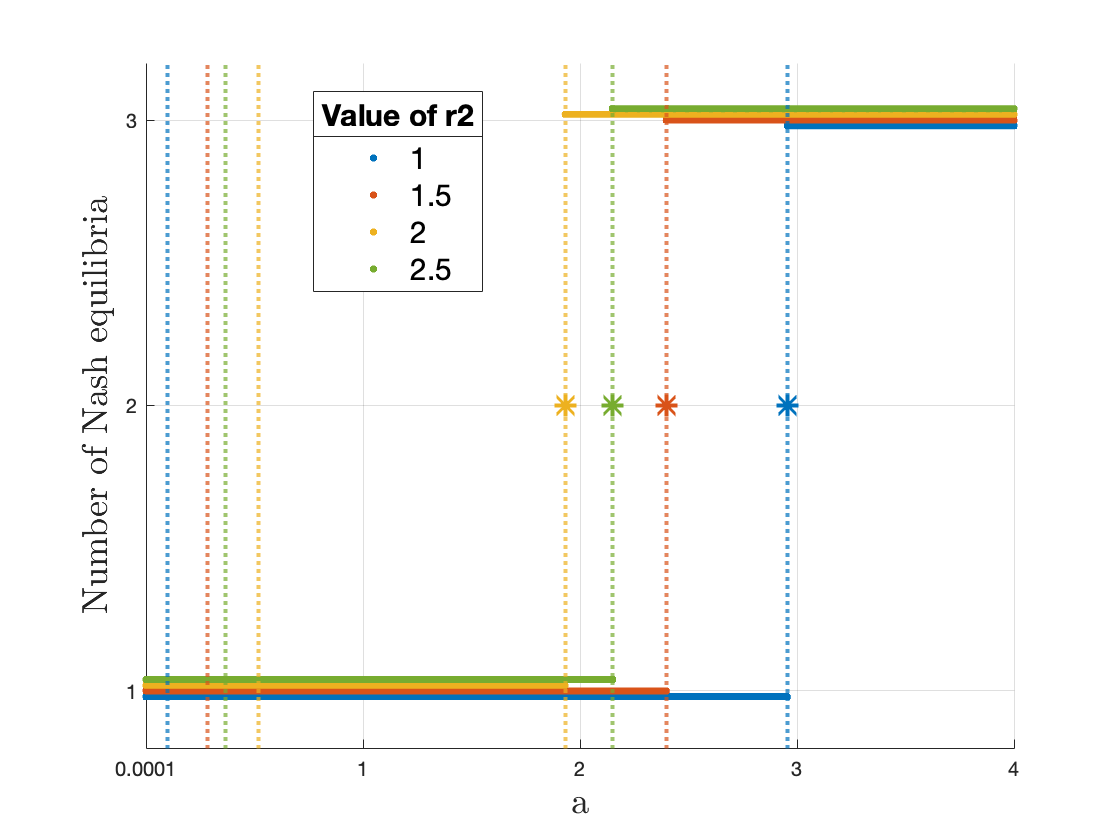}
    \end{subfigure}
    \caption{\textbf{Top:} The discriminant as a function of $a$ for different values of $r_2$, the $y-$axis is in symmetric log scale. \textbf{Bottom:} The Number of Nash equilibria under the same conditions.}
    \label{fig:DandNE}
\end{figure}

We consider a game with $q_1 = 0.5$, $r_1 = 1$, and $q_2 = 1$. We evaluate $\Delta_g$ as a function of the system parameter $a$ for $a$ in the range $[0.0001,4]$, and for four values of $r_2$. Then, we compute the number of Nash equilibria for the system by solving equation \eqref{eq:Grobner} and verifying that they are also solutions of the system \eqref{sys:derivatives}. In Figure \ref{fig:DandNE}, we plot the curves, corresponding to the four values of $r_2$,  of the discriminant (top figure) and the number of Nash equilibria (bottom figure) as a function of $a$. The vertical dotted lines indicate the roots of $\Delta_g$. We found two such roots for each value of $r_2$, and we denote the smallest one as $a_{min,r_2}$ and the largest one as $a_{max,r_2}$.
We observe that for all values of $r_2$, there is a unique Nash equilibrium for $a_{min,r_2}$ and two Nash equilibria for $a_{max,r_2}$. These results are consistent with Statements 1 and 3 of Theorem \ref{thm:main}. Additionally, there are two regions where the discriminant is positive. In the first regions, where $a \in (0, a_{min,r_2})$, there exists a unique Nash equilibrium. In the second region, where $a \in (a_{max,r_2}, 4]$; the system possesses three Nash equilibria. Hence, Statements 1 and 2 of Theorem \ref{thm:main} are respected. Finally, notice that the discriminant is negative for $a \in (a_{min,r_2}, a_{max,r_2})$ and the Nash equilibrium is unique, as predicted by Statements 4 in Theorem \ref{thm:main}.

\vspace{-0.1cm}
\section{CONCLUSION}

We investigated the existence, computation, and number of Nash equilibria for scalar discrete-time infinite-horizon LQ games with two agents. Our results proved the existence of a Nash equilibrium, determined the maximum number of Nash equilibria, and provided a quintic polynomial whose real roots include all Nash equilibria. Our findings were based on leveraging the Gröbner basis concept from \cite{buchberger1985grobner}. This algebraic geometry concept allowed us to establish that the considered LQ games possess at most three Nash equilibria, and to provide sufficient conditions for the number of Nash equilibria to be at most two and sufficient conditions for the uniqueness. We also conducted numerical experiments illustrating cases where the number of Nash equilibria is equal or strictly smaller than the upper bounds we provided. 




\section*{ACKNOWLEDGMENT}
\vspace{-0.1cm}
Reda Ouhamma thanks Yassine El Maazouz for the helpful discussions on algebraic techniques and Macaulay2.  Giulio Salizzoni is financially supported by the Swiss National Science Foundation, grant number 207984. 


\vspace{-0.1cm}
\bibliographystyle{IEEEtran}
\bibliography{ref}

\addtolength{\textheight}{-12cm}   

\end{document}